\documentclass[9pt]{article}
\usepackage{fullpage}
\usepackage{authblk}
\usepackage{etoolbox}
\usepackage[numbers,sort&compress,square,comma]{natbib}
\usepackage[colorlinks,citecolor=blue]{hyperref}
\makeatletter
\pretocmd{\NAT@open}{\begingroup\color{\@citecolor}}{}{}
\apptocmd{\NAT@close}{\endgroup}{}{}
\makeatother
\usepackage{chngpage}
\usepackage{amsfonts}
\usepackage{amsthm}
\usepackage{amsmath}
\usepackage{amssymb}
\usepackage{algorithm}
\usepackage{algorithmicx}
\usepackage{subfig}
\usepackage{graphicx}
\usepackage{color}
\usepackage[section]{placeins}
\usepackage{bbm}
\usepackage[noend]{algpseudocode}
\usepackage{booktabs}

\def\w{{\bf w}}

\def\x{{\bf x}}
\def\y{{\bf y}}

\def\1{{\bf{1}}}
\def\0{{\bf{0}}}

\def\R{\mathbb{R}}

\def\1{{\bf{1}}}
\def\0{{\bf{0}}}

\def\w{{\bf w}}

\def\x{{\bf x}}
\def\y{{\bf y}}

\title{A Scalable Algorithm for Structure Identification of Complex Gene Regulatory Network from Temporal Expression Data}

\author[1]{Shupeng Gui}
\author[2]{Rui Chen}
\author[3]{Liang Wu}
\author[1,4*]{Ji Liu}
\author[3*]{Hongyu Miao}
\affil[1]{Department of Computer Science, University of Rochester, Rochester, 14620, USA}
\affil[2]{Molecular and Human Genetics, Baylor College of Medicine, Houston, 77030, USA}
\affil[3]{Department of Biostatistics, School of Public Health, UTHealth, Houston, 77030, USA}
\affil[4]{Goergen Institute for Data Science, University of Rochester, Rochester, 14620, USA}

\date{\today}

\begin{document}
\maketitle
\thispagestyle{plain} \pagestyle{plain}

\begin{abstract}
\noindent \textbf{Motivation:} Gene regulatory interactions are of fundamental importance to various biological functions and processes. However, only a few previous computational studies have claimed success in revealing genome-wide regulatory landscapes from temporal gene expression data, especially for complex eukaryotes like human. Moreover, recent work suggests that these methods still suffer from the curse of dimensionality if network size increases to 100 or higher. \\

\noindent \textbf{Result:} We present a novel scalable algorithm for identifying genome-wide regulatory network structures. The highlight of our method is that its superior performance does not degenerate even for a network size on the order of $10^4$, and is thus readily applicable to large-scale complex networks. Such a breakthrough is achieved by considering both prior biological knowledge and multiple topological properties (i.e., sparsity and hub gene structure) of complex networks in the regularized formulation. We also illustrate the application of our algorithm in practice using the time-course expression data from an influenza infection study in respiratory epithelial cells. \\

\noindent \textbf{Availability and Implementation:} The algorithm described in this article is implemented in MATLAB$^\circledR$. The source code is freely available from \url{https://github.com/Hongyu-Miao/DMI.git}. \\

\noindent \textbf{Contact:} jliu@cs.rochester.edu; hongyu.miao@uth.tmc.edu \\

\noindent \textbf{Supplementary information:} Supplementary data are available online.
\end{abstract}

\section{Introduction}
Gene regulatory network (GRN), consisting of multiple regulators and their target molecules, plays critical roles in numerous biological processes by modulating the expression levels of RNAs and proteins \citep{Karlebach2008}. While remarkable successes in dissecting single genes that are responsible for certain biological functions, behavior or diseases have been achieved over the past few decades, it has been increasingly recognized that elucidating gene functions and interactions in the context of networks becomes more and more important to gain novel insight into mechanisms, effects and interventions of molecular, cellular or organ-level biological processes \citep{Barabasi2004, Barabasi2011, Kidd2014}. Clearly, one of the prerequisites for investigators to harvest the benefits of such systematic network approaches is whether the structures of gene regulatory networks can be accurately revealed from experimental data.

Modern high-throughput experimental technologies such as next generation sequencing \citep{Metzker2010} can generate time-course data at a much more affordable cost \citep{Chen2013}, thus provide unprecedented opportunities for researchers to systematically investigate the temporal patterns of gene expressions and infer gene regulatory relationships. However, two well-known major obstacles have significantly hampered our ability to interrogate such data for novel scientific findings. First, limited by resources or technical and ethic issues, the sampling frequency of time-course gene expression profiling data is low (e.g., most of the time-course GEO datasets \citep{Edgar2002} have less than 6 time points), which renders the sample size far less than the number of unknown parameters in the context of GRN structure identification. Targeting at such scenarios, it is of significant importance to borrow information from additional sources (e.g., previous biological knowledge). Second, considering the fact that for complex eukaryotes like human, the number of protein-coding genes is approximately {\color{black}19,000 \citep{Ezkurdia2014}} so the problem dimension is {\it ultra-high} (i.e., tens of thousands or even millions unknown parameters are involved). The development of novel and more efficient algorithms that can scale to such high-dimensional networks is still necessary.

{\color{black}A number of modeling and computational approaches have been developed for gene network structure identification \citep{Marbach2012}, including information theory method \citep[e.g.][]{OpgenRhein2007}, clustering method \citep[e.g.][]{Langfelder2008}, Boolean network \citep{Shmulevich2002}, Bayesian network \citep[e.g.][]{Hartemink2006}, state-space model \citep{Hirose2008}, regression model \citep{Yeung2002}, and differential equation model \citep[e.g.][]{DeJong2002}. The well-known methods for static expression data include the Pearson correlation coefficient (PCC) and the mutual information (MI) methods \citep{Bansal2007, MarbachPrill2010}, the ARACNE algorithm \cite{margolin2006aracne} based on the data processing inequality concept \citep{CoverThomas1991}, the context likelihood of relatedness (CLR) approach \cite{Faith2007large} that extends the relevance networks method \citep{ButteKohane2000}, the GENIE3 algorithm \cite{HuynhThu2010} that uses tree-based gene ranking and ensemble, and the TIGRESS algorithm \cite{Haury2012tigress} that combines the least angle regression with stability selection. However, only a few algorithms were specifically developed for time-course data. \cite{Zoppoli2010} proposed to calculate the time-delayed dependencies between gene expressions using mutual information, and developed and tested the TimeDelay-ARACNE algoithm on networks with less than 30 genes. Our group introduced the concept of background network to confine the parameter space using prior biological knowledge and built the SITPR pipeline based on a dynamic Bayesian network (DBN) model regularized by $L^1$ norm \citep{Liu2014SITPR}. More recently, \cite{Huynh-Thu2015} developed the Jump3 method by combining the on/off stochastic differential equation model with decision trees, where the marginal likelihood of each node was used as the decision function.}

{\color{black}The DREAM Challenge \citep{Schaffter2011} makes it feasible to generate authoritative benchmark data for algorithm performance evaluation.} To the best knowledge of our authors, none of the algorithms above have been evaluated on DREAM networks of a size greater than 100; actually, their performances all become unsatisfactory at a network size of 100, as verified in several previous studies \citep{Bansal2007, MarbachPrill2010, Liu2014SITPR}. Therefore, a breakthrough in algorithm development is necessarily needed to obtain accurate and reliable results for large-scale  GRN inference (e.g., a network size O($10^4$)). The work of \cite{Liu2014SITPR} is one of the few studies that systematically borrow information from previous biological knowledge to deal with the low sampling frequency and the high dimensionality issues in GRN inference; however, even after the incorporation of prior knowledge, the number of unknown parameters is still on the order of $10^5$ so the high dimensionality issue remains. Since GRNs are typical complex networks with structural sparsity \citep{Leclerc2008}, it is common to impose sparsity on the inference problem formulation in previous studies \citep{Yeung2002, Haury2012tigress, Liu2014SITPR}; however, it turned out that sparsity alone cannot satisfyingly address the high dimensionality problem \citep{Liu2014SITPR}. This observation leads us to the hypothesis that considering additional structural or topological properties of large-scale complex networks may result in novel scalable algorithms for GRN inference that are not subject to the curse of dimensionality.

In this study, we adopt the background network approach developed in our previous study \citep{Liu2014SITPR} for parameter space confinement, and we modify other selected state-of-the-art algorithms to take the advantage of the same background network for fairness of algorithm performance comparison. More importantly, a breakthrough in algorithm performance is achieved by considering both structural sparsity and the existence of hub genes, which is a prominent topological feature of GRNs due to preferential gene attachment \citep{Barabasi2004}. We describe the mathematical formulation and computational steps of the novel algorithm in Section \ref{SectMethods}. We evaluate and compare the performance of our algorithm with other state-of-the-art approaches using DREAM4 benchmark data that are generated from networks of a size up to $2 \times 10^4$, and we apply the proposed method to real temporal gene expression data from an influenza H1N1 virus infection study to illustrate its usage in practice. The related computational settings and results are presented in Section \ref{sec:results}. Finally, we discuss the advantages and limitations of this work in Section \ref{SectDiscuss}.

\begin{figure*} [t]
\centering
\includegraphics[width=165mm]{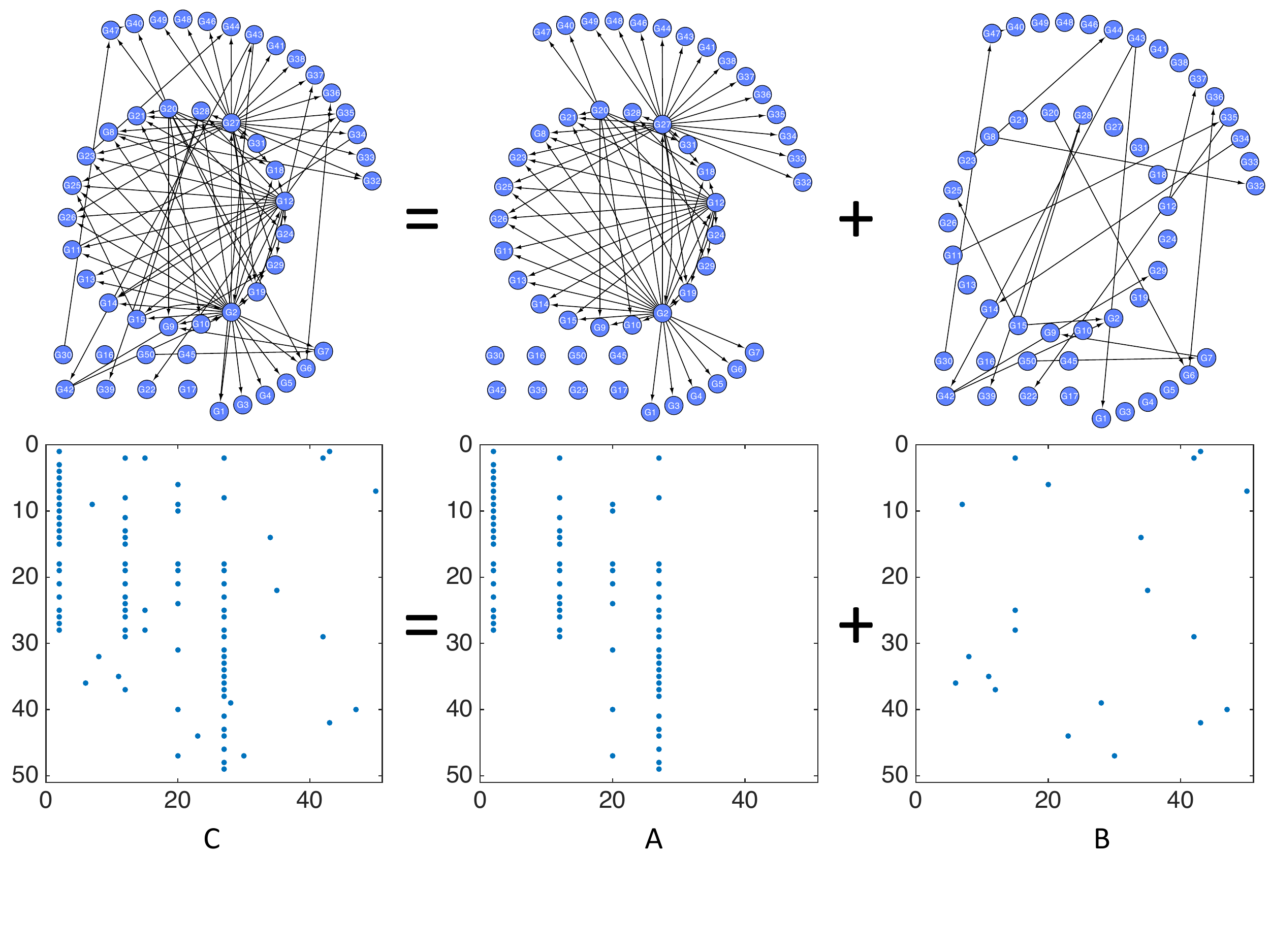}
\vspace{-10mm}
\caption{Illustration of the hub gene structure separation and the corresponding coefficient matrices.}
\label{fig:sgsaddition}
\end{figure*}

\section{Methods}\label{SectMethods}
\subsection{Background regulatory network} \label{sec:bkgnetwork}
For ultra-high dimensional problems considered in this study, it is important to constrain the parameter space based on prior biological information to improve computing efficiency and avoid over-fitting, especially for datasets with limited sample sizes. For this purpose, the concept of background regulatory network was introduced in our previous work \citep{Liu2014SITPR}, and we recently also developed a curated database called RegNetwork for public use \citep{Liu2015regnetwork}. Here we present the key features of the background network and describe how it is employed in this study for GRN inference.

First, the background regulatory network is comprehensive so it contains not only experimentally-observed regulatory interactions but also physically, chemically or biologically feasible interactions. For this purpose, information from over 20 selected databases or knowledgebases (e.g., FANTOM \citep{Ravasi2010}, TRANSFAC \citep{Matys2006}, JASPAR \citep{Bryne2008}, KEGG \citep{Kanehisa2000}, HPRD \citep{Mishra2006}, and {\color{black}ENCODE \citep{Gerstein2012})} have been collected and integrated. In addition, potential interactions between transcription factors and target genes are predicted based on sequence motifs obtained from Ensembl \citep{Flicek2012}. In this way, the background network also allows the discovery of novel regulatory relationships that have not been officially reported or documented. Second, the background network is not cell type- or condition-specific but it allows the detection of cell type- or condition-specific regulatory relationships. The reason is that under different conditions (e.g., different diseases), only certain regulatory interactions in certain types of cells will be activated in the background network, and an appropriately designed algorithm should be able to detect a reasonable number of activated interactions with the presence of the inactive ones. Third, the background network is different from random networks \citep{Barabasi1999} in terms of certain properties like characteristic path length and node degree distribution \citep{Albert2005}; also, its node degrees are found to satisfy the power-law distributions so the background network is scale-free.

Here we use the background regulatory network concept in both the simulation studies and the real data example. That is, our primary task is to identify the activated interactions in the confined parameter space specified by the background network. We recognize that some existing algorithms under comparison start with the full network structure; therefore, for fairness, we make necessary efforts to modify their computer codes to take the background network as the parameter space. Also, it should be mentioned that the same weight is assigned to all the edges in the background network in this study for simplicity. That means, our algorithm will automatically determine the sparse structure and the hub gene structures without informing itself the potential location of such structures beforehand. The algorithm performance can be further improved if edge weights can be assigned based on evidence strength in, e.g., literature; however, even with a simple equal-weight scheme, our algorithm can already achieve a satisfying performance (see the Results Section). Finally, for simulation studies, the background network is introduced by adding additional random edges to the DREAM4 network structures; for real data example, the background network for human fetched from the RegNetwork database is used, which contains 23,079 nodes and 372,774 edges.

\subsection{Model formulation of topological features of GRN} \label{sec:formulation}
While sparsity of GRNs has been extensively explored in previous studies, the hub gene structure has rarely been addressed in existing formulations of GRN inference. This study suggests that in addition to sparsity, the hub gene structure is also of significant importance to the development of scalable algorithms that are not subject to the curse of dimensionality.

Throughout this paper, the following notations are used:
\begin{itemize}
	\item $n$ denotes the total number of nodes in a network;
	\item $\x_t \in \mathbb{R}^{n}$ denotes the expression level of $n$ genes (and miRNAs) at time $t$;
	\item $\odot$ denotes the Hadamard product (i.e., the product of the entries of two matrices at the same position). For example, $[1, 3] \odot [0,2] = [0, 6]$.
	\item $\|\cdot\|_\mathcal{F}$ is the Frobenius norm of a matrix $M$: $\|M\|_\mathcal{F} = \sqrt{\sum_{i,j}M^2_{ij}}$;
    \item $\|.\|$ is the Spectral norm (i.e., the maximal singular value of a matrix);
    \item $C_{\cdot j}$ : $j$th column of $C$;
    \item $C_{i\cdot}$ : $i$th row of $C$.
\end{itemize}

The dynamic Bayesian network (DBN) model \citep{Zou2005DBN} is considered in this study for GRN inference. It has been shown by previous studies \citep{Kim2003DBN, Zou2005DBN} that under the normality and Markov assumptions, the DBN model can be reduced to a vector autoregression (VAR) model as follows:
 \begin{equation} \label{eq:dynsys}
 \x_{t+1} = P \x_t + \w_t,\quad t=1,\cdots, T-1,
 \end{equation}
where $T$ is the total number of time points, and $P\in \R^{n\times n}$ is the coefficient matrix, characterizing the change from time $t$ to the next time point $t+1$. $P_{ij} \neq 0$ indicates the existence of regulatory relationship between regulator $j$ and target $i$. {\color{black}Also, for simplicity and robustness, uneven time intervals are treated as of equal length.} We consider an equivalent form of Eq. \eqref{eq:dynsys}:
\begin{equation}\label{eq:diff_dynsys}
	\y_{t+1,t} := \x_{t+1}-\x_t = C \x_t + \w_t,
\end{equation}
where $C = P-I$. Our goal is to estimate the coefficient matrix $C$ (or equivalently $P$)  given the time-course observations $\{\x_t\}_{t=1}^{T}$. The objective function can be formulated as follows
\begin{equation}
\min_{C}\quad {1\over 2} \|Y - CX\|^2_\mathcal{F} = \sum_{t=1}^{T-1} \|(\x_{t+1}-\x_t) - C \x_t\|^2,
\label{eq:loss}
\end{equation}
where $Y := [\x_2-\x_1, \x_3-\x_2, \cdots, \x_{T} - \x_{T-1}] \in \R^{n\times (T-1)}$ and $X := [\x_1, \x_2, \cdots, \x_{T-1}] \in \R^{n\times (T-1)}$.

Since the total number of distinct time points are usually small, the solution obtained by directly solving \eqref{eq:loss} suffers from the overfitting issue. To overcome this problem, the commonly used strategy is to confine the model space using regularizations or constraints; for example, the $\ell_1$ norm regularization for imposing model sparsity. However, in addition to sparsity, there exist other prominent topological characteristics for complex networks like GRNs {\color{black}\citep{Costa2007}}. To illustrate this, we use the Yeast gene network as an example. The top left graph in Fig.~\ref{fig:sgsaddition} shows a typical substructure of the Yeast gene network; the bottom left figure is the corresponding coefficient matrix $C$, where the blue dots indicate nonzeros (3.6\%) and the white areas indicate zeros (96.4\%) in $C$. Sparsity is thus a clear feature to tell from Fig.~\ref{fig:sgsaddition}, and we also consider the following network characteristics in our model formulation.

\paragraph{\bf Hub gene structure separation}: As suggested in Fig.~\ref{fig:sgsaddition}, there exist a small number of gene nodes with a high degree, called the ``hub'' genes. If we define the {\it hub gene structure} as the set of outgoing edges from a hub gene node to its immediate neighbors, we can separate this structure from the original GRN such that matrix $C$ can be decomposed into
\[
C = A + B,
\]
where matrix $A$ corresponds to the hub gene structure (the upper middle graph in Fig.~\ref{fig:sgsaddition}) and matrix $B$ corresponds to the remaining edges (the upper right graph in Fig.~\ref{fig:sgsaddition}).

It can be told from the lower middle plot of Fig.~\ref{fig:sgsaddition} that matrix $A$ for hub gene structure has a column group sparse structure (that is, it only contains a few nonzero columns), while the matrix $B$ for the remaining edges shows a more uniform sparsity pattern. Therefore, we consider the $\ell_{2,1}$ norm convex regularization on $A$ to enforce the group sparsity structure \citep{huang2010benefit},
\[
\|A\|_{2,1}:=\sum_j (\sum_i |A_{ij}|^2)^{\frac{1}{2}},
\]
and consider the commonly used $\ell_{1}$ regularization \citep{candes2005l1} on $B$ to enforce the uniform sparsity,
\[
\|B\|_1:= \sum_{i,j} |B_{ij}|.
\]

\paragraph{\bf Small number of parent nodes}: Based on the real GRN structures previously reported (e.g., Yeast), we find that while one gene may directly modulate the expressions of many other genes, a single gene is usually co-regulated by only a few regulators. That is, in a GRN represented by a directed acyclic graph, the number of parent nodes of a node is often small. This observation suggests the non-zero elements in any row of matrix $C$ should be small (see the bottom left of Fig.~\ref{fig:sgsaddition}). We thus can impose the following constraint on the rows of matrix $C$
\[
\|C_{i\cdot}\|_0 \leq \xi_i, \quad i=1,\cdots, n,
\]
where $\xi_i$ is a predefined upper bound of the number of incoming edges to gene $i$ (e.g., {\color{black}0.7 times the indegree of a node in the background network}).

\paragraph{\bf Background network}: The {\it mask of the background network} can be denoted as $\Omega \in \{0, 1\}^{n\times n}$, where $\Omega_{ij}=1$ corresponds to the non-existence of any regulatory interaction between genes $i$ and $j$ and $\Omega_{ij}=0$ corresponds to the existence of a possible interaction between genes $i$ and $j$ that needs to be determined from temporal expression data. Therefore, the background network constraints imposed on matrix $C$ can be denoted as follows:
\[
\mathcal{P}_{\Omega}(C):= C \odot \Omega = 0.
\]

Combining all the constraints above, the overall GRN structure identification formulation is given below:
\begin{equation}
\begin{aligned}
	\min_{A,B, C} \quad& \frac{1}{2}\|(A+B)X-Y\|_{\mathcal F}^2 + \alpha \|A\|_{2,1} +\beta\|B\|_1\\
	\textrm{s.t.}\quad& \|C_{i\cdot}\|_0 \leq \xi_i\\
				  & \mathcal{P}_\Omega(A) = 0,
                   \mathcal{P}_\Omega(B) = 0\\
				  & A+B = C.
\end{aligned}
\label{eq:objADMM}
\end{equation}
where $\alpha$ and $\beta$ are two penalty coefficients that balance the weights on group sparsity and uniform sparsity.

\subsection{Computing algorithm} \label{sec:computing}
Existing solvers cannot deal with the proposed formulation above, so we introduce an efficient optimization algorithm, called Decomposable Multi-structure Identification (DMI) (Algorithm \ref{alg:DMI}), that can handle a network size greater than $20k$ on an average computer \citep{boyd2011distributed}.

We first define the Augmented Lagrangian function:
\begin{equation}
\begin{aligned}
	& L(A,B,C,U) = \frac{1}{2}\|(A+B)X-Y\|_\mathcal{F}^2 + \alpha\|A\|_{2,1}+ \beta\|B\|_1\\
                & \quad +\frac{\rho}{2}\|A+B-C\|_{\mathcal{F}}^2+ \langle U,A+B-C\rangle\\
                & \quad + \mathbb{I}_{\|C_{i\cdot}\|_0\leq\xi_i}(C) + \mathbb{I}_{\mathcal{P}_{\Omega}(A) = 0}(A) + \mathbb{I}_{\mathcal{P}_{\Omega}(B) = 0}(B),
\end{aligned}
\label{eq:LagABCU}
\end{equation}
where $\rho$ is a positive real number, $\mathbb{I}_{\text{condition}}(\cdot)$ denotes the function that gives 0 if the argument variable satisfies the condition and gives positive infinity otherwise, {\color{black}and $U$ is a dual matrix of the same size as $C$ introduced to solve the optimization problem}.

\begin{algorithm}[htb]
\caption{Decomposable Multi-structure Identification (DMI)}
\label{alg:DMI}
\begin{algorithmic}
\Require $X$, $Y$ , $\xi_i, \rho > 0$
\Ensure $C$
\Repeat
	\State $[A,B] = \textrm{arg}\min_{A,B}\quad L(A,B,C,U)$ using Algorithm \ref{alg:PGDAB};
    \State $C = \textrm{arg}\min_{C} \quad L(A,B,C,U)$ to update $C$ using \eqref{eq:updateC};
	\State Update $U$ by $U \leftarrow U + \rho (A+B-C)$;
\Until convergence;\\
\Return $C$.
\end{algorithmic}
\end{algorithm}

More specifically, we can break the optimization problem into minimization w.r.t. $A$ and $B$, minimization w.r.t. $C$, and {\color{black}minization} w.r.t. $U$ as follows.

\paragraph{\bf Minimization w.r.t. $A$ and $B$.}
Since some terms in Eq. \eqref{eq:LagABCU} such as $\mathbb{I}_{\|C_{i\cdot}\|_0\leq\xi_i}(C)$ contain no $A$ or $B$, the optimization problem can be simplified:
\begin{equation}
\begin{aligned}
	\min_{A,B} \quad &\frac{1}{2}\|(A+B)X-Y\|_\mathcal{F}^2 + \alpha\|A\|_{2,1} + \beta\|B\|_1\\
    &+\frac{\rho}{2}\|A+B-C\|_{\mathcal{F}}^2 + \langle U,A+B-C\rangle \\
    &+ \mathbb{I}_{\mathcal{P}_{\Omega}(A) = 0}(A) + \mathbb{I}_{\mathcal{P}_{\Omega}(B) = 0}(B),
\end{aligned}
\label{eq:objAB}
\end{equation}
which can be solved using the coordinate block descent method \citep{tseng2001convergence}. By taking partial derivatives of the objective function, we can derive
\begin{equation}
	g = [(A+B)X-Y]X^\top + \rho (A+B-C) + U;\notag
\end{equation}
and by applying the proximal gradient descent method \citep{nesterov2005smooth}, the following update rules for $A$ and $B$ at each iteration can be obtained
\begin{align*}
A_{\cdot i} &=\textrm{prox}_{\gamma\alpha\|\cdot\|}((A - \gamma g_A)_{\cdot i})\\
	&=\max\Big(0,1-\frac{\gamma\alpha}{\|(A - \gamma g_A)_{\cdot i}\|}\Big)(A - \gamma g_A)_{\cdot i},
\end{align*}
and
\begin{align*}
B &= \textrm{prox}_{\gamma\beta\|\cdot\|_1}(B - \gamma g_B)\\
&= \textrm{sign}(B - \gamma g_B) \odot \max(0,|B - \gamma g_B|-\gamma\beta),
\end{align*}
where $\textrm{sign}(a) = 1$ if $a>0$, $\textrm{sign}(a) = 0$ if $a = 0$, and otherwise $-1$. Usually we choose $\gamma = \frac{1}{\|X\|^2 + \rho}$ as the safe step length for each update to guarantee the monotonic decreasing of the objective function \eqref{eq:objAB}. More details about the accelerated proximal gradient descent method are given in Algorithm \ref{alg:PGDAB}.

\paragraph{\bf Minimization w.r.t. $C$.}
Similar to {\color{black}the objective function} \eqref{eq:objAB}, some terms in {\color{black}Eq.} \eqref{eq:LagABCU} like $\mathbb{I}_{\mathcal{P}_{\Omega}(A) = 0}(A)$ do not contain matrix $C$. Hence,  {\color{black}Eq.} \eqref{eq:LagABCU} can be simplified as {\color{black}follows}:
\vspace{-3mm}
\begin{equation}
\begin{aligned}
	\min_C \quad & \frac{\rho}{2}\|A+B-C\|_\mathcal{F}^2 + \langle U,A+B-C\rangle + \mathbb{I}_{\|C_{i\cdot}\|_0\leq \xi_i}(C),\\
	\propto& \frac{1}{2}\|A+B-C+\frac{1}{\rho}U\|_\mathcal{F}^2 + \mathbb{I}_{\|C_{i\cdot}\|_0\leq \xi_i}(C).
\end{aligned}
\label{eq:objC}
\end{equation}
\vspace{-5mm}

\hspace{-4mm}We thus update $C$ by the following rule
\vspace{-3mm}
\begin{equation}
	\begin{aligned}
		C = \mathcal{P}_{\xi}(A+B+\frac{1}{\rho}U),
	\end{aligned}
    \label{eq:updateC}
\end{equation}
\vspace{-5mm}

\hspace{-4mm}where $\mathcal{P}_{\xi}$ is a projection of $\|C_{i\cdot}\|_0\leq \xi_i$ by retaining the largest $\xi_i$ elements in the $i$-th row of $C$.

\paragraph{\bf {\color{black}Minimization} w.r.t. $U$.}
Similar to the derivation of {\color{black}the objective function} \eqref{eq:objC} above, we have
\vspace{-3mm}
\begin{equation}
\begin{aligned}
	\min_U \quad & \langle U,A+B-C\rangle.
\end{aligned}
\label{eq:objU}
\end{equation}
\vspace{-5mm}

We update $U$ using $U = U + \rho (A+B-C)$ for each iteration.

\begin{algorithm}[tb]
\caption{Accelerated Proximal Gradient Descent Method \cite{nesterov1983method} }
\label{alg:PGDAB}
\begin{algorithmic}
\Require n (number of genes in network), $\gamma$ (step length), $\alpha$, $\beta$, $C$, $U$
\Ensure $A$, $B$
\State Initialize $t_1 =1 $, $V_1 = A_1 = \0$, $W_1 = B_1 = \0$, $\rho = 10$;
\For {k = 1: K}
	\State $g_k = \mathcal{P}_{\Omega}([(V_k+W_k)X-Y]X^\top + \rho (V_k+W_k-C) + U)$;
	\For {i=1 to n } \State{${A_{k+1}}_{\cdot i} =\textrm{prox}_{\gamma\alpha\|\cdot\|}((V_k - \gamma g_k)_{\cdot i})$;
} \EndFor
	\State  $B_{k+1} =  \textrm{prox}_{\gamma\beta\|\cdot\|_1}(W_k - \gamma g_k)$;	
	\State $t_{k+1} = \frac{1}{2}+\frac{1}{2}\sqrt{1+4t_k^2}$;
	\State $V_{k+1} = A_{k+1} +  (\frac{t_{k}-1}{t_{k+1}})(A_{k+1}-A_k)$;
	\State $W_{k+1} = B_{k+1} +  (\frac{t_{k}-1}{t_{k+1}})(B_{k+1}-B_k)$;
\EndFor
\Return $A$ and $B$.
\end{algorithmic}
\end{algorithm}

\subsection{Algorithmic complexity}
Let $|\bar{\Omega}|$ denote the total number of zeros in the background network matrix $\Omega \in \mathbb{R}^{n\times n}$ used in Algorithms \ref{alg:DMI} and \ref{alg:PGDAB}. Algorithm \ref{alg:DMI} is assumed to run $J$ iterations to converge, and for each iteration of Algorithm \ref{alg:DMI}, we run Algorithm \ref{alg:PGDAB} for $K$ iterations. Therefore, the complexity of Algorithm \ref{alg:PGDAB} is $O(|\bar{\Omega}|\times T\times K)$, where $X,Y\in \mathbb{R}^{n\times (T-1)}$, $T \ll n$ and $|\bar{\Omega}| \ll n^2$. Therefore, the complexity of DMI is $O(|\bar{\Omega}|\times T\times K\times J)$.

\enlargethispage{6pt}

\section{Results and discussion} \label{sec:results}
\subsection{Evaluation and comparison}
\subsubsection{Network specification and simulated data}
To evaluate and compare the proposed method with other state-of-the-art algorithms, we employ GeneNetWeaver \citep{Marbach2009DREAM3, Schaffter2011}, the official DREAM Challenge tool for time-course expression data generation. More specifically, GeneNetWeaver uses the GRNs from Yeast (4,441 nodes, 12,873 edges) or Ecoli (1,565 nodes, 3,785 edges) as templates to generate network structures with typical complex network properties, of a network size up to $\sim4,400$; and then ordinary differential equation (ODE) models are built upon the previously generated network structures to produce time-course gene expression data at pre-specified time points.

In our simulation studies, we use GeneNetWeaver to generate the network structures for a given network size $n=10, 100, 1000$, or $20,000$, and we treat such networks as the ground truth. Let $m$ denote the number of edges in a ground truth network, we then randomly add $m$ additional edges to the ground truth network to generate the background network with $2m$ edges. Consequently, the goal of GRN inference algorithms in the simulation studies is to identify as many ground truth edges as possible from all the background network edges, with a controlled false positive rate. It should be stressed that networks of a size $20,000$ cannot be generated from the Yeast or Ecoli templates, so we supply the large-scale GRNs from the RegNetwork database \citep{Liu2015regnetwork} to GeneNetWeaver as templates (e.g., the human GRN in RegNetwork has $\sim23,000$ nodes and $\sim370,000$ edges). For each simulated dataset, one data point is generated at each of 6 distinct time points, respectively, for a pre-specified noise level to match the observation scheme in the real data example in Section \ref{sec:realdata}.

\subsubsection{Fairness of comparison and evaluation metrics}
Since the state-of-the-art algorithms under comparison such as ARACNE \citep{margolin2006aracne} start with a full matrix $C$, for fairness of comparison, we make efforts for these existing algorithms to also take the advantage of the background network {\color{black}by confining the calculation of the performance evaluation criteria to the background network}. The computing parameters used by existing algorithms are tuned as suggested in their original manuscripts {\color{black}(see Table S1 in Supplementary Data I)}. For our DMI algorithm, we set $\alpha = 0.08$, $\beta = 0.16$ and $\rho = 10$ for all the simulated datasets,{\color{black}where $\alpha$ and $\beta$ are determined using cross-validation}.

Five commonly-used criteria are considered to measure algorithm performance, including sensitivity ($SN$), specificity ($SP$), accuracy ($ACC$), Matthews correlation coefficient ($MCC$), and the Area Under ROC Curve ($AUC$):
\vspace{-3mm}
\begin{align*}
	SN &= \frac{TP}{TP+FN},\\
	SP &= \frac{TN}{TN + FP},\\
	ACC &= \frac{TP+\color{black}{TN}}{TP+FP+TN+FN},\\
	MCC &=\frac{TP\times TN - FP\times FN}{\sqrt{(TP+FP)(TP+FN)(TN+FP)(TN+FN)}},
\end{align*}
where $TP$ and $TN$ denote the true positive and true negative, and $FP$ and $FN$ denote the false positive and false negative, respectively. \\


\begin{table*}[ht]
\begin{adjustwidth}{-.5in}{-.5in}
\begin{center}
\caption{Performance evaluation of DMI and other competing algorithms on a network size 10, 100, or 1,000 at a 10\% noise level.}
\label{tab:diff_method}
\scriptsize
\begin{tabular}{llllllll}
\hline
Methods & Size & SN                    & SP                    & ACC                               & MCC                   & AUC                   \\
\hline
DMI     & 10   & \bf{0.7600 $\pm$0.2119}    & 0.9143 $\pm$0.1380    & \bf{0.8235 $\pm$0.1301}         & \bf{0.6837 $\pm$0.2180}    & \bf{0.8371 $\pm$0.1187}    \\
        & 100  & \bf{0.7321 $\pm$0.1818}    & 0.8807 $\pm$0.0401    & \bf{0.8069 $\pm$0.0825}         & \bf{0.6264 $\pm$0.1460}    & \bf{0.8064 $\pm$0.0831}    \\
        & 1000 & \bf{0.8119 $\pm$0.0472}    & 0.9023 $\pm$0.0147    & \bf{0.8564 $\pm$0.0245}        & \bf{0.7171 $\pm$0.0460}    & \bf{0.8571 $\pm$0.0242}    \\
CLR     & 10   & 0.5900$\pm$0.1776     & 0.4143$\pm$0.3125     & 0.5176$\pm$0.1331           & 0.0043$\pm$0.3125     & 0.5457$\pm$0.1331     \\
        & 100  & 0.4854$\pm$0.0200     & 0.4916$\pm$0.0400     & 0.4885$\pm$0.0202           & -0.0231$\pm$0.04      & 0.4768$\pm$0.0202     \\
        & 1000 & 0.5063$\pm$0.0134     & 0.4908$\pm$0.0269     & 0.4987$\pm$0.0135           & -0.0029$\pm$0.0269    & 0.4292$\pm$0.0135     \\
PCC & 10   & 0.5600$\pm$0.0938 & 0.3714$\pm$0.1698 & 0.4824$\pm$0.0719   & -0.0686$\pm$0.1698 & 0.5386$\pm$0.0719 \\
    & 100  & 0.5244$\pm$0.0268 & 0.5301$\pm$0.0536 & 0.5273$\pm$0.0268   & 0.0545$\pm$0.0536  & 0.5148$\pm$0.0268 \\
    & 1000 & 0.5061$\pm$0.0172 & 0.4907$\pm$0.0344 & 0.4985$\pm$0.0173   & -0.0032$\pm$0.0344 & 0.4290$\pm$0.0173 \\
MINET   & 10   & 0.5500$\pm$0.0834     & 0.5571$\pm$0.1745     & 0.5529$\pm$0.0915           & 0.1076$\pm$0.1745     & 0.5871$\pm$0.0915     \\
        & 100  & 0.3378$\pm$0.0319     & 0.6988$\pm$0.0351     & 0.5194$\pm$0.0183           & 0.0391$\pm$0.0351     & 0.5073$\pm$0.0183     \\
        & 1000 & 0.2464$\pm$0.0164     & 0.7801$\pm$0.0196     & 0.5092$\pm$0.0096           & 0.0313$\pm$0.0196     & 0.4224$\pm$0.0096    \\
TIGRESS & 10   & 0.5700$\pm$0.0949 & 0.3857$\pm$0.1355 & 0.4941$\pm$0.1116   & -0.0443$\pm$0.2304 & 0.5171$\pm$0.0865 \\
        & 100  & 0.3154$\pm$0.0239 & 0.6904$\pm$0.0405 & 0.5040$\pm$0.0276   & 0.0067$\pm$0.0595  & 0.4898$\pm$0.0281 \\
        & 1000 & 0.0062$\pm$0.0007 & 0.9941$\pm$0.0014 & 0.4926$\pm$0.0006   & 0.0024$\pm$0.0077  & 0.3630$\pm$0.0008 \\
ARACNE  & 10   & 0.2800$\pm$0.1200     & 0.7286$\pm$0.1966     & 0.4647$\pm$0.0962           & 0.0074$\pm$0.1966     & 0.6043$\pm$0.0962     \\
        & 100  & 0.0598$\pm$0.0325     & 0.9514$\pm$0.0357     & 0.5083$\pm$0.0103          & 0.0225$\pm$0.0357     & 0.4860$\pm$0.0103     \\
        & 1000 & 0.0188$\pm$0.0094     & 0.9820$\pm$0.0246     & 0.4930$\pm$0.0040           & 0.0023$\pm$0.0246     & 0.3670$\pm$0.0040     \\
TimeDelay-ARACNE  & 10  & 0.0500$\pm$0.1214  & \textbf{0.9571$\pm$0.2029} & 0.4235$\pm$0.0744            & 0.0222$\pm$0.2029  & 0.6386$\pm$0.0744 \\
        & 100  & 0.0118$\pm$0.0250 & \textbf{0.9912$\pm$0.0596}  & 0.5044$\pm$0.0073           & 0.0000$\pm$0.0596  & 0.4793$\pm$0.0073  \\
        & 1000 & 0.0034$\pm$0.0028  & \textbf{0.9968$\pm$0.0179}  & 0.4924$\pm$0.0015           & 0.0011$\pm$0.0179  & 0.3622$\pm$0.0015   \\
GENIE3           & 10   & 0.5600$\pm$0.0138         & 0.3714$\pm$0.0419            & 0.4824$\pm$0.0098         & -0.0686$\pm$0.0419        & 0.5400$\pm$0.0098         \\
                 & 100  & 0.5093$\pm$0.0009         & 0.5153$\pm$0.0035            & 0.5123$\pm$0.0010         & 0.0246$\pm$0.0035         & 0.5007$\pm$0.0010         \\
                 & 1000 & 0.5075$\pm$0.0001         & 0.4922$\pm$0.0003            & 0.5000$\pm$0.0001         & -0.0003$\pm$0.0003        & 0.4304$\pm$0.0001         \\
Jump3  & 10     & 0.6400$\pm$0.1095    &0.4857$\pm$0.2048      & 0.5765$\pm$0.1094     & 0.1257$\pm$0.2048   & 0.6000$\pm$0.1094  \\
                    & 100    & 0.2606$\pm$0.0431    & 0.8390$\pm$0.0474     & 0.5515$\pm$0.0245       & 0.1217$\pm$0.0474   & 0.5471$\pm$0.0245  \\
        & 1000  & 0.0384$\pm$0.0074        & 0.9843$\pm$0.0192        &  0.5040$\pm$0.0035               &  0.0694$\pm$0.0192       &  0.3823$\pm$0.0035   \\
SITPR   & 10   & 0.4900$\pm$0.1197     & 0.7143$\pm$0.2020     & 0.5824$\pm$0.0757           & 0.2172$\pm$0.1819     & 0.6021$\pm$0.0862     \\
        & 100  & 0.1610$\pm$0.0794     & 0.8253$\pm$0.0656     & 0.4952$\pm$0.0255           & -0.0212$\pm$0.0733    & 0.4931$\pm$0.0257     \\
        & 1000 & 0.2382$\pm$0.0509     & 0.7599$\pm$0.0381     & 0.4950$\pm$0.0118           & -0.0032$\pm$0.0279    & 0.4990$\pm$0.0114     \\
        \hline
\end{tabular}
\end{center}
\end{adjustwidth}
\end{table*}



\begin{table*}[h]
\centering
\caption{Evaluation of DMI at a noise level from 10\% to 30\% for a network size 10, 100, 1,000 or 20,000.}
\label{tab:diff_noise}
\scriptsize
\begin{tabular}{llllllll}
\hline
Noise Level & Size & SN                  & SP                  & ACC                 & MCC                 & AUC                 \\ \hline
10\% noise  & 10   & 0.9333 $\pm$ 0.0573 & 0.9000 $\pm$ 0.0860 & 0.9200 $\pm$ 0.0688 & 0.8333 $\pm$ 0.1434 & 0.9166 $\pm$ 0.0717 \\
            & 100  & 0.7687 $\pm$ 0.0379 & 0.8253 $\pm$ 0.0407 & 0.7984 $\pm$ 0.0283 & 0.5965 $\pm$ 0.0571 & 0.7970 $\pm$ 0.0282 \\
            & 1000 & 0.8273 $\pm$ 0.0244 & 0.8305 $\pm$ 0.0131 & 0.8289 $\pm$ 0.0182 & 0.6579 $\pm$ 0.0363 & 0.8289 $\pm$ 0.0182 \\
            & 20k  & 0.7748 $\pm$ 0.0041 & 0.7753 $\pm$ 0.0041 & 0.7751 $\pm$ 0.0041 & 0.5502 $\pm$ 0.0083 & 0.7751 $\pm$ 0.0041\\
20\% noise  & 10   & 0.8000 $\pm$ 0.1365 & 0.7000 $\pm$ 0.2048 & 0.7600 $\pm$ 0.1639 & 0.5000 $\pm$ 0.3414 & 0.7500 $\pm$ 0.1707 \\
            & 100  & 0.7312 $\pm$ 0.0440 & 0.7569 $\pm$ 0.0398 & 0.7447 $\pm$ 0.0418 & 0.4881 $\pm$ 0.0838 & 0.7440 $\pm$ 0.0419 \\
            & 1000 & 0.8056 $\pm$ 0.0133 & 0.8006 $\pm$ 0.0137 & 0.8031 $\pm$ 0.0135 & 0.6063 $\pm$ 0.0271 & 0.8031 $\pm$ 0.0135 \\
            & 20k  & 0.7452 $\pm$ 0.0045 & 0.7457 $\pm$ 0.0045 & 0.7454 $\pm$ 0.0045 & 0.4909 $\pm$ 0.0091 & 0.7454 $\pm$ 0.0045\\
30\% noise  & 10   & 0.7333 $\pm$ 0.1194 & 0.6000 $\pm$ 0.1791 & 0.6800 $\pm$ 0.1433 & 0.3333 $\pm$ 0.2986 & 0.6667 $\pm$ 0.1493 \\
            & 100  & 0.7005 $\pm$ 0.0320 & 0.7291 $\pm$ 0.0289 & 0.7155 $\pm$ 0.0304 & 0.4297 $\pm$ 0.0610 & 0.7148 $\pm$ 0.0305 \\
            & 1000 & 0.7829 $\pm$ 0.0080 & 0.7774 $\pm$ 0.0082 & 0.7802 $\pm$ 0.0081 & 0.5603 $\pm$ 0.0162 & 0.7801 $\pm$ 0.0081 \\
            & 20k  & 0.7339 $\pm$ 0.0048 & 0.7344 $\pm$ 0.0048 & 0.7342 $\pm$ 0.0048 & 0.4684 $\pm$ 0.0096 & 0.7342 $\pm$ 0.0048 \\\hline
\end{tabular}
\end{table*}

\begin{figure*}[ht]
\centering
\includegraphics[width=\textwidth]{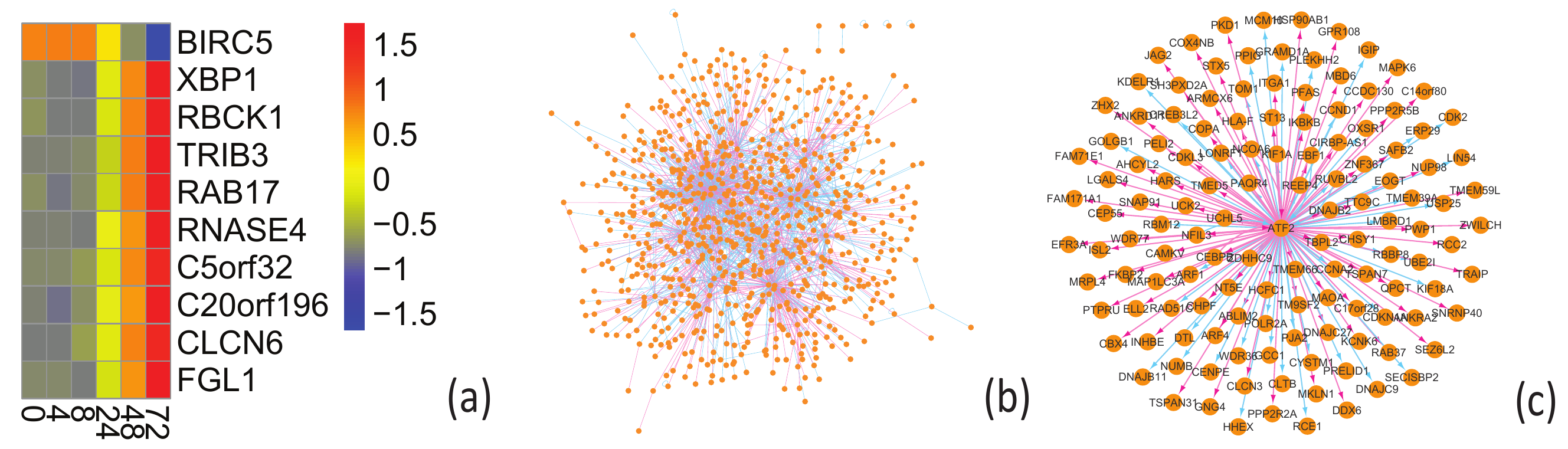}
\vspace{-5mm}
\caption{Application of the DMI algorithm to the expression data from human A549 cells in response to influenza H1N1 virus infection. (a) Example of differentially expressed genes{\color{black},where the color bar values are the normalized gene expression levels}; (b) Overall GRN structure {\color{black}(the full details can be found on GitHub)}; (c) ATF2 hub gene structure.}
\label{fig:A549GRN}
\end{figure*}


\newpage

\subsubsection{Performance evaluation}
The first set of experiments are conducted to compare our DMI algorithm with nine representative algorithms, including PCC, ARACNE \citep{margolin2006aracne}, CLR \citep{Faith2007large}, MINET {\color{black}(the maximum relevance minimum redundancy method)} \citep{Meyer2008MINET}, GENIE3 \citep{HuynhThu2010}, {\color{black}TimeDelay-ARACNE \citep{Zoppoli2010}}, TIGRESS \citep{Haury2012tigress}, SITPR \citep{Liu2014SITPR}, {\color{black}and Jump3 \citep{Huynh-Thu2015}}. Specifically, SITPR \citep{Liu2014SITPR} is a multi-step pipeline with the $\ell_1$ norm penalty incorporated; to avoid manual intervention needed by certain SITPR steps, we simply compare the constrained LASSO step in SITPR with DMI. In the simulated data, the noise level is fixed at 10\%, and the networks size ranges from 10 to 1,000 {\color{black}because other competing algorithms cannot handle a network size of $20,000$}.

For the second set of experiments, the noise level is increased from $10\%$ to $30\%$ to evaluate the robustness of our DMI algorithm against noise. The network size in the corresponding simulated data is 10, 100, 1000 or 20,000.

The two sets of experiment results are summarized in Tables \ref{tab:diff_method} and \ref{tab:diff_noise}, respectively. From Table \ref{tab:diff_method}, it can be clearly told that for network size ranging from 10 to 1,000, our method significantly outperform other state-of-the-art methods; e.g., the AUCs of our method remain 0.8 or higher for all cases, but the AUCs of all other methods are 0.6 or less and they drop to $0.36\sim0.50$ for $n=1,000$. From Table \ref{tab:diff_noise}, one can tell that our DMI algorithm is robust against noise because the performance decrease (measured by AUC) is less than 0.04 as the noise level increases from 10\% to 30\%, even for a network size of 20,000. {\color{black}Additional experiment results can be found in Supplementary Data I.}

\subsection{Regulatory network identification in human respiratory epithelial cells during IAV infection} \label{sec:realdata}
The real data example for illustrating the use of DMI in practice is from the recent study of \cite{Loveday2012}, where human A549 cells were infected with influenza H1N1 virus (A/Mexico/InDRE4487/2009). Illumina HumanHT-12 v3 BeadChips and Febit miRBase 14 Geniom miRNA Biochips were employed to measure the mRNA and miRNA samples, respectively. Six replicates of the expression levels were collected at each of six time points (0, 4, 8, 24, 48, and 72 hours post infection), but one sample at hour 4 was excluded due to degradation. The within-sample normalized dataset is available from NCBI GEO (GSE36553 and GSE36461). We further conducted between-sample normalization on the data across the six time points.

The original analyses in \cite{Loveday2012} did not use all the time points; instead, a small number of regulatory relationships between functional miRNAs and gene targets were inferred based on data on hr 0 and hr 8 only. The genome-wide regulatory landscape has not been revealed based on the entire dataset although \cite{Liu2014SITPR} made an attempt to investigate the subnetwork structures of a size from 2 to $300$. In this study, 1,572 genes and 14 non-coding RNAs were identified to be differentially expressed (Figure \ref{fig:A549GRN}(a)) using the FPCA method \citep{Wu2013} for a false discovery rate controlled at 0.05. From the human background network in RegNetwork, the DMI algorithm was applied to identify 1,926 regulatory interactions between the differentially expressed genes or miRNAs, as shown in Figure \ref{fig:A549GRN}(b).

We conducted comprehensive network analyses on the topological structure and found that the A549 GRN in response to influenza H1N1 infection is a typical complex network, evidenced by, e.g., its power law degree distribution and a large clustering coefficient (see Supplementary Data I for details). A close examination of the inferred GRN by the DMI algorithm suggests consistency between the findings in this study and those in literature. For instance, 'CEBPB' (CCAAT/Enhancer-Binding Protein Beta) is a transcription factor (TF) that plays an important role in immune and inflammatory responses \citep{Rebhan1998}, and the subnetwork structure centered at 'CEBPB' has been previously reported in \cite{Liu2014SITPR}. Our study also revealed a similar subnetwork structure around 'CEBPB' (e.g., interactions between 'CEBPB' and 'NCS1', 'ISG20', or 'ABCG1'. See Figure S2 in Supplementary Data I); however, some experimentally-verified interactions (e.g., between 'CEBPB' and 'NOLC1' in HPRD) were successfully identified in this study, but were missing in the work of \cite{Liu2014SITPR}. More interestingly, we found a number of hub gene structures, e.g., centered at the Activating Transcription Factor genes ('ATF2' and 'ATF4') or the E2F Transcription Factor genes ('E2F6' and 'E2F7'). More specifically, 'ATF2' encodes a DNA binding protein of the leucine zipper family that can perform distinct functions via different mechanisms (e.g., formation of a homo/heterodimer with 'c-Jun'). Our results showed that 'ATF2' interacts with more than 130 other genes, including 'CEBPB' (Figure \ref{fig:A549GRN}(c)). Although it has been previously reported that influenza A viral RNA molecules can indirectly activate transcription factors like ATF2 \citep{Kochs2007}, the importance of ATF2 in the context of influenza infection has not been fully appreciated by previous experimental studies, considering the large number of target genes associated with ATF2 in the hub gene structure. The proposed DMI algorithm thus provides us not only an opportunity to investigate the genome-wide regulatory landscapes but also a way to identify interesting subnetworks like hub gene structures. {\color{black}Further experiments like ChiP-PCR for measuring gene-TF interactions can be conducted in the future to validate the results produced by DMI.}

\section{Conclusions} \label{SectDiscuss}
We proposed a novel scalable algorithm called DMI for large-scale GRN inference from temporal gene expression data in this study. Extensive simulation studies and real data application suggest the superiority of the DMI algorithm over other state-of-the-art approaches, and the success mainly relies on the incorporation of multiple topological characteristics of GRNs like sparsity and hub gene structures into our model formulation.

We also recognize several limitations of the current study: 1) DMI's performance can be further improved if the background network edges can be appropriately weighted; 2) Prediction of previously unseen interactions heavily depends on the background network preparation so further efforts need to be invested to the development of public knowledgebases like RegNetwork; 3) This work mainly focuses on solving the ultra-high dimensional inference problems, and data heterogeneity and integration issues have not been tackled. We believe that it is necessary to address the aforementioned limitations in separate articles, and this study provides a solid basis for such future investigations.

\vspace*{-10pt}

\section*{Acknowledgements}

We thank Dr. Zhiping Liu (Shandong University, China) and Dr. Hulin Wu (UTSPH) for developing the RegNetwork database. We also thank Dr. Andrew P. Rice (BCM) for constructive feedbacks. \vspace*{-12pt}

\section*{Funding}

This work was partially supported by NIH/NEI grants R01EY022356, R01EY018571, and R01EY020540 (RC), NSF grant CNS-1548078 (JL), {\color{black}and NSF grant DMS-1620957 (HM)}. \vspace*{-12pt}

\bibliographystyle{abbrv}

\bibliography{document}

\end{document}